\documentclass[aps,twocolumn,amsmath,amssymb]{revtex4}
\usepackage{times}
\usepackage{microtype}
\usepackage{graphicx}
\usepackage{amsfonts}
\usepackage{amsmath}
\usepackage{amsthm}
\usepackage{amssymb}
\usepackage{dsfont}
\usepackage{color}
\usepackage{soul}
\usepackage{bm}
\usepackage{ulem}

\definecolor{DarkRed}{rgb}{0.65,0,0}
\definecolor{DarkBlue}{rgb}{0,0,0.65}

\begin{document}

\title{Tunability of Andreev levels via spin-orbit coupling in Zeeman-split Josephson junctions}
\author{Tatsuki Hashimoto$^{1,2}$, Alexander A. Golubov$^{1,3}$, Yukio Tanaka$^2$, and Jacob Linder$^{4,5}$}
\affiliation{$^1$Faculty of Science and Technology and MESA+ Institute of Nanotechnology, University of Twente, 7500 AE, Enschede, The Netherlands}
\affiliation{$^2$Department of Applied Physics, Nagoya University, Nagoya 464-8603, Japan}
\affiliation{$^3$Moscow Institute of Physics and Technology, Dolgoprudny, Moscow 141700, Russia}
\affiliation{$^4$Department of Physics, NTNU, Norwegian University of Science and Technology, N-7491 Trondheim, Norway}
\affiliation{$^5$Center of Excellence QuSpin, NTNU, Norwegian University of Science and Technology, N-7491 Trondheim, Norway}
\date{\today}
\begin{abstract}
We study Andreev reflection and Andreev levels $\varepsilon$ in Zeeman-split superconductor/Rashba wire/Zeeman-split superconductor junctions by solving the Bogoliubov de-Gennes equation. We theoretically demonstrate that the Andreev levels $\varepsilon$ can be controlled by tuning either the strength of Rashba spin-orbit interaction or the relative direction of the Rashba spin-orbit interaction and the Zeeman field. In particular, it is found that the magnitude of the band splitting is tunable by the strength of the Rashba spin-orbit interaction and the rength of the wire, which can be interpreted by a spin precession in the Rashba wire. We also find that if the Zeeman field in the superconductor has the component parallel to the direction of the junction, the $\varepsilon$-$\phi$ curve becomes asymmetric with respect to the superconducting phase difference $\phi$. Whereas the Andreev reflection processes associated with each pseudospin band are sensitive to the relative orientation of the spin-orbit field and the exchange field, the total electric conductance interestingly remains invariant.
\end{abstract}

\maketitle
\section{Introduction}
The Josephson effect is the fundamental phenomenon in superconductor junctions \cite{JOSEPHSON1962251}. 
Since the discovery of this effect, various types of structure have been studied. In particular, the superconductor/ferromagnet/superconductor (S/F/S) junction has  attracted much research interest because of its high tunability of the supercurrent \cite{RevModPhys.76.411,RevModPhys.77.1321,RevModPhys.77.935}.
In S/F/S junctions, the so called $\pi$ phase, where the direction of the critical current is reversed compared with 0 phase, is realized by changing the strength of the exchange field or thickness of the ferromagnetic region \cite{ryazanov_prl_01,Bulaevskii}. 

In recent years, superconductors with spin-split energy bands, so called Zeeman-split superconductors (ZSs), have also been studied widely owing to their potential application to the superconducting spintronics \cite{JLJR_review,0034-4885-78-10-104501}. The homogeneous spin splitting in the superconductor can be realized in the systems such as thin F/S junctions \cite{PhysRevLett.86.3140,Golubov2002} or thin superconductor films under the application of an in-plane magnetic field \cite{MESERVEY1994173}. It has been shown that N/ZS junctions, where N stands for a normal metal, can generate highly spin-polarized current \cite{PhysRevB.77.132501,PhysRevB.78.014516,Emamipour201617,PhysRevB.75.132503}. Josephson junctions with spin-split superconductors have been also studied in various types of structures. In ZS/N/ZS junctions, the spin degeneracy of the Andreev level is lifted and the magnitude of the spin splitting can be controlled by changing either the magnitude or relative direction of the Zeeman field in the superconducting leads \cite{0295-5075-115-6-67001,PhysRevB.65.134507,1009-1963-16-11-060,0253-6102-52-4-32}. ZS/N/ZS junctions with unconventional superconducting pairing such as $p$- or $d$-wave pairing, have also been studied and the tunability of the Andreev level by changing the relative direction of the Zeeman field and superconducting $d$-vector have been demonstrated \cite{Emamipour201411,1674-1056-23-5-057402,Emamipour2014,0253-6102-43-3-035}.

\begin{figure}[t]
\begin{center}
\includegraphics[width=8cm]{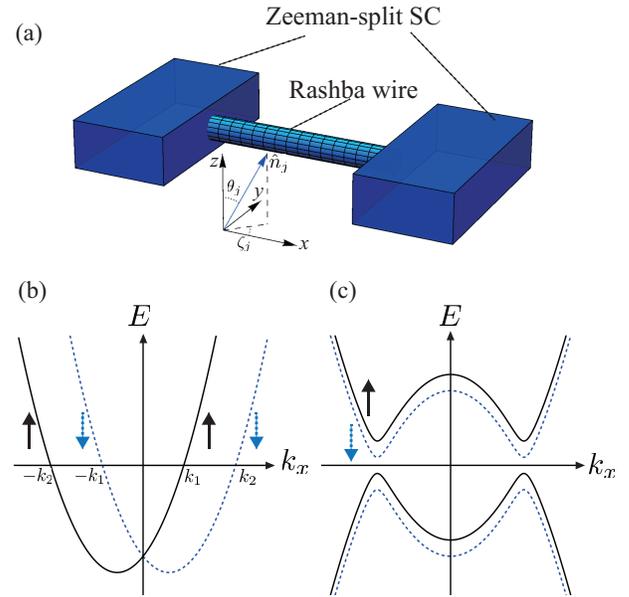}
\caption{(a) Schematic illustration of the Zeeman-split superconductor/Rashba wire/Zeeman-split superconductor junction. $\hat{n}_{\eta}=(\cos\theta_\eta \cos \zeta_\eta,\cos\theta_\eta \sin \zeta_\eta,\sin\theta_\eta)$ is an unit vector representing the direction of Zeeman field in the left ($\eta=L$) and right ($\eta=R$) superconductor. (b) Schematic band structure of a Rashba wire. (c) Schematic band structure of a Zeeman-split superconductor. Black solid (blue dotted) lines show the spin-up (spin-down) band.}
\label{fig_setup}
\end{center}
\end{figure}

The magnetic tunability of the Andreev levels and the resulting supercurrent has been shown in the previous studies as mentioned above. However, the electric tunability of these levels has not been discovered yet in the ZS junctions, even though the electric tunability tends to have advantages for nano-device applications. To realize the electric tunability, the most promising way is introducing the Rashba spin-orbit interaction (RSOI) in a system, which is tunable by the gate voltage \cite{doi:10.1063/1.102730,RSOI}. Recently, there has been a growing interest in a one-dimensional Rashba wire, especially after the proposal to use it as a platform for Majorana fermions \cite{PhysRevLett.103.020401,PhysRevLett.105.077001,PhysRevLett.105.177002}. The Rashba wire has a characteristic band structure, where the spin degeneracy is lifted and the direction of spin and momentum is locked (spin-momentum locking). More importantly, because of the pseudomagnetic effect of the RSOI, the spin precession takes place while an electron or hole is traveling in the Rashba wire. 

Another motivation for this work is that although the effect of the exchange field in the nanowire is often taken into account in recent literature considering the prospect of Majorana fermions in spin-orbit coupled Josephson junctions with externally applied magnetic field, the fact that even a very small exchange field $h$ induced in the superconducting region (in the case where these are thin enough to permit this) may affect the physical properties of the system remains virtually unexplored in this context. Despite the fact that a nanowire made of a material such as InAs is likely to have a higher $g$-factor than the materials used for the superconducting regions, even a small exchange field $h \ll \Delta$ induced in the superconductors is sufficient to induce qualitatively new physics such as considerable thermoelectric effects \cite{machon_prl_13, ozaeta_prl_14, linder_prb_16}. Our study is therefore also of relevance with regard to Majorana experiments utilizing sufficiently thin superconducting regions that an in-plane field may induce a small exchange splitting in them.

In this paper, we theoretically study Andreev reflection and the formation of Andreev levels in Zeeman-split superconductor/Rashba wire/ Zeeman-split superconductor (ZS/RW/ZS) junction to reveal how the electric tunable RSOI in the wire and the Zeeman field in the superconductors affects the Andreev level. We first study Andreev and normal reflections at the boundary of the RW/ZS bilayer, since these processes are of fundamental relevance to the formation of bound states in a Josephson geometry. We then calculate the Andreev level energies in the ZS/RW/ZS Josephson junction. We find that the Andreev levels can be controlled by the strength of the RSOI $\lambda$, the length of the Rashba wire $l$ and the direction of the Zeeman field though the tunneling conductance is not so affected by the RSOI. It is found that the magnitude of the band splitting of the Andreev level oscillates as a function of the strength of the RSOI $\lambda$ and the length of the Rashba metal $l$ with a certain period. We also find that the Andreev level can be dramatically altered by changing the direction of the Zeeman field relative to the vector characterizing the RSOI. In particular, if the Zeeman field has a component along the junction ($x$ component in this paper), the Andreev level becomes asymmetric in the superconducting phase difference $\phi$ and provides a finite supercurrent even at zero phase-bias $\phi=0$. 

This paper is organized as follows. In Sec.\ref{sec_form}, we first introduce a model Hamiltonian for a junction consisting of the ZS and Rashba wire. In Sec.\ref{sec_AR}, calculation results of the Andreev reflection and the tunneling conductance are discussed.
In Sec.\ref{sec_AL_noRSOI}, we review the Andreev level in the absence of RSOI. Then we move to the main results: Andreev level in the presence of RSOI in Sec.\ref{sec_AL_RSOI}. Finally, we summarize our results in Sec.\ref{sec_summary}.

%%%%%%%%%%%%%%%%%%%%%%%%%%%%%%%%%%%%%%%%%%%%%%%%%%%%%%%%%%%%%%%%%%%
%%%%%%%%%%%%%%%%%%%%%%%%%%%%%%%%%%%%%%%%%%%%%%%%%%%%%%%%%%%%%%%%%%%
%%%%%%%%%%%%%%%%%%%%%%%%%%%%%%%%%%%%%%%%%%%%%%%%%%%%%%%%%%%%%%%%%%%
%%%%%%%%%%%%%%%%%%%%%%%%%%%%%%%%%%%%%%%%%%%%%%%%%%%%%%%%%%%%%%%%%%%
\section{Formulation}\label{sec_form}
Figure \ref{fig_setup}(a) shows the schematic picture of the ZS/RW/ZS junction. In this paper, we consider a short ballistic junction that satisfies $l\ll \xi$, where $l$ is the length of the Rashba wire and $\xi$ is the ballistic superconducting coherence length.
The Bogoliubov-de Gennes (BdG) Hamiltonian for this system is described by,
\begin{align}
\begin{pmatrix}
\hat{H}_0(r)&\hat{\Delta}(x)\\
-\hat{\Delta}^*(x)&-\hat{H}_0^*(r)\\
\end{pmatrix}
\hat{\psi}_i(r)
=\varepsilon\hat{\psi}_i(r),\label{eq_BdG}
\end{align}
with
\begin{align}
\hat{H}_0(r)&=\left[-\frac{\hbar^2}{2m}\nabla^2-\mu+Z(x)\right]\hat{\sigma}_0-\hat{\lambda}(x)-\hat{h}(x)\label{eq_RSOI},\\
Z(x)&=Z[\delta(x)+\delta(x-l)],\\
\hat{\lambda}(x)&=
\left\{k_x,\lambda\Theta(x)\Theta(l-x)\right\}\hat{\sigma}_y\\
\hat{h}(x)&=h_L\Theta(-x)\hat{\bm n}_{\bm L}\cdot\hat{\bm \sigma}+h_R\Theta(x-l)\hat{\bm n}_{\bm R}\cdot\hat{\bm \sigma},\\
\hat{\Delta}(x)&=\Delta[e^{i\phi_L}\Theta(-x)+e^{i\phi_R}\Theta(x-l)](i\hat{\sigma}_y),
\end{align}
where the basis is set as ($c_\uparrow,c_\downarrow,c_\uparrow^\dagger,c_\downarrow^\dagger$) and $\bm \hat \sigma$ is the Pauli matrix for the spin space. 
Here, $Z(x)$ denotes the barrier potential at the boundaries, $\hat{\lambda}(x)$ is the Rashba spin-orbit interaction in the normal region, $\hat{h}(x)$ is the exchange field in the superconducting region and $\hat{\Delta}(x)$ is the superconducting pair potential, where $\delta(x)$ and $\Theta(x)$ are the $\delta$ function and step function, respectively. To satisfy the Hermiticity of the Hamiltonian at the boundaries, we adopt $\left\{k_x,\lambda\Theta(x)\Theta(l-x)\right\}\hat{\sigma}_y$ as a Rashba spin-orbit interaction term instead of $\lambda k_x\hat{\sigma}_y$.
In addition, $\hat{\bm n}_{\bm \eta}=(\cos\theta_\eta \cos \zeta_\eta,\cos\theta_\eta \sin \zeta_\eta,\sin\theta_\eta)$ is a unit vector representing the direction of the Zeeman field in the left ($\eta=L,$) and right ($\eta=R$) superconductors. In this paper, we focus on the conventional $s$-wave superconductivity, hence we assume that the magnitude of the superconducting gap $\Delta$ is constant and positive value. The band structure of the Rashba wire and Zeeman-split superconductor are shown in Figs.\ref{fig_setup} (b) and (c), where the systems are considered as infinite.

By diagonalizing the model Hamiltonian, one can obtain the wave function in the superconducting region under the plane-wave assumption as
\begin{align}
\hat{\psi}_{L}(x)&=
a_{L}e^{-iq^{+}_{\uparrow}x}
[
u_1\cos \frac{\theta_L}{2}e^{i\phi_L/2},
u_1\sin \frac{\theta_L}{2}e^{i(\zeta_L+\phi_L/2)},
\nonumber
\\
&\hspace{3pc}
-v_1\sin \frac{\theta_L}{2}e^{i(\zeta_L-\phi_L/2)},
v_1 \cos \frac{\theta_L}{2}e^{-i\phi_L/2}
]^T
\nonumber\\
&+b_{L}e^{-iq^{+}_{\downarrow}x}
[
u_2\sin \frac{\theta_L}{2}e^{-i(\zeta_L-\phi_L/2)},
-u_2\cos \frac{\theta_L}{2}e^{i\phi_L/2},
\nonumber
\\
&\hspace{3pc}
v_2\cos \frac{\theta_L}{2}e^{-i\phi_L/2},
v_2 \sin \frac{\theta_L}{2}e^{-i(\zeta_L+\phi_L/2)}
]^T
\nonumber\\
&+c_{L}e^{iq^{-}_{\uparrow}x}
[
v_1\cos \frac{\theta_L}{2}e^{i\phi_L/2},
v_1\sin \frac{\theta_L}{2}e^{i(\zeta_L+\phi_L/2)},
\nonumber
\\
&\hspace{3pc}
-u_1\sin \frac{\theta_L}{2}e^{i(\zeta_L-\phi_L/2)},
u_1\cos \frac{\theta_L}{2}e^{-i\phi_L/2}
]^T
\nonumber\\
&+d_{L}e^{iq^{-}_{\downarrow}x}
[
v_2\sin \frac{\theta_L}{2}e^{-i(\zeta_L-\phi_L/2)},
-v_2\cos \frac{\theta_L}{2}e^{i\phi_L/2},
\nonumber
\\
&\hspace{3pc}
u_2\cos \frac{\theta_L}{2}e^{-i\phi_L/2},
u_2\sin \frac{\theta_L}{2}e^{-i(\zeta_L+\phi_L/2)}
]^T,
\end{align}
\begin{align}
\hat{\psi}_{R}(x)&=
a_{R}e^{iq^{+}_{\uparrow}x}
[
u_1\cos \frac{\theta_R}{2}e^{i\phi_R/2},
u_1\sin \frac{\theta_R}{2}e^{i(\zeta_R+\phi_R/2)},
\nonumber
\\
&\hspace{3pc}
-v_1\sin \frac{\theta_R}{2}e^{i(\zeta_R-\phi_R/2)},
v_1 \cos \frac{\theta_R}{2}e^{-i\phi_R/2}
]^T
\nonumber\\
&+b_{R}e^{iq^{+}_{\downarrow}x}
[
u_2\sin \frac{\theta_R}{2}e^{-i(\zeta_R-\phi_R/2)},
-u_2\cos \frac{\theta_R}{2}e^{i\phi_R/2},
\nonumber
\\
&\hspace{3pc}
v_2\cos \frac{\theta_R}{2}e^{-i\phi_R/2},
v_2 \sin \frac{\theta_R}{2}e^{-i(\zeta_R+\phi_R/2)}
]^T
\nonumber\\
&+c_{R}e^{-iq^{-}_{\uparrow}x}
[
v_1\cos \frac{\theta_R}{2}e^{i\phi_R/2},
v_1\sin \frac{\theta_R}{2}e^{i(\zeta_R+\phi_R/2)},
\nonumber
\\
&\hspace{3pc}
-u_1\sin \frac{\theta_R}{2}e^{i(\zeta_R-\phi_R/2)},
u_1\cos \frac{\theta_R}{2}e^{-i\phi_R/2}
]^T
\nonumber\\
&+d_{R}e^{-iq^{-}_{\downarrow}x}
[
v_2\sin \frac{\theta_R}{2}e^{-i(\zeta_R-\phi_R/2)},
-v_2\cos \frac{\theta_R}{2}e^{i\phi_R/2},
\nonumber
\\
&\hspace{3pc}
u_2\cos \frac{\theta_R}{2}e^{-i\phi_R/2},
u_2\sin \frac{\theta_R}{2}e^{-i(\zeta_R+\phi_R/2)}
]^T,
\end{align}
where $a_j$, $b_j$, $c_j$ and $d_j$ with $j=L, R$ are the coefficients of electron-like quasiparticles with spin-$\uparrow$, hole-like quasiparticles with spin-$\uparrow$, electron-like quasiparticles with spin-$\downarrow$, and hole like quasiparticles with spin-$\downarrow$, respectively. Moreover, we define
\begin{align}
u_{1(2)}&=
\sqrt{
\frac{1}{2}
\left(
1+\frac{\sqrt{[\varepsilon+(-)h]^2-\Delta^2}}{\varepsilon+(-)h}
\right)}
\\
v_{1(2)}&=
\sqrt{
\frac{1}{2}
\left(
1-\frac{\sqrt{[\varepsilon+(-)h]^2-\Delta^2}}{\varepsilon+(-)h}
\right)}.
\end{align}
The wave vectors $q^\pm_{\uparrow,\downarrow}$ are represented as
\begin{align}
q_{\uparrow,\downarrow}^\pm&=\sqrt{(2m/\hbar^2)(\varepsilon_F^{S}\pm\Omega_{\uparrow,\downarrow})},\\
\Omega_{\uparrow(\downarrow)}&=\sqrt{[\varepsilon+(-) h]^2-\Delta^2}.
\end{align}
We assume $\varepsilon_F^S\gg |\Omega|$ so that the wave-vectors can be treated as $q_\uparrow^+=q_\downarrow^+=q_\uparrow^-=q_\downarrow^-\equiv q_F$.
In a similar manner, the total wave-function in the normal region is described by 
\begin{align}
\hat{\psi}_N(x)&=\frac{1}{\sqrt{2}}
[
a_1e^{ik_{1}x}
(i,1,0,0)^T
+a_2e^{ik_{2}x}
(-i,1,0,0)^T
\nonumber
\\
&+
b_1e^{-ik_{1}x}
(-i,1,0,0)^T
+b_2e^{-ik_{2}x}
(i,1,0,0)^T
\nonumber
\\
&+c_1e^{-ik_{1}x}
(0,0,-i,1)^T
+c_2e^{-ik_{2}x}
(0,0,i,1)^T
\nonumber
\\
&+d_1e^{ik_{1}x}
(0,0,i,1)^T
+d_2e^{ik_{2}x}
(0,0,-i,1)^T],
\label{eq_wave_RW}
\end{align}
with
\begin{align}
k_{1}&=-\lambda+\sqrt{\lambda^2+k_F^2},\\
k_{2}&=\lambda+\sqrt{\lambda^2+k_F^2},
\end{align}
where $a_j$, $b_j$, $c_j$ and $d_j$ are coefficients of a right-moving electron, a left-moving electron, a right-moving hole, and a left-moving hole with wave number $k_j$ ($j=1,2$).
Here, we set $\hbar=m=1$ for brevity. The boundary conditions for the wave functions are given by
\begin{align}
\hat{\psi}_L(0)-\hat{\psi}_N(0)&=0,\\
\hat{\psi}_N(l)-\hat{\psi}_R(l)&=0,\\
\left.\frac{\partial \hat{\psi}_L(x)}{\partial x}\right|_{x=0}
-\left.\frac{\partial \hat{\psi}_N(x)}{\partial x}\right|_{x=0}
&=(Z\hat{I}+\lambda\hat{\tau})\hat{\psi}_N(0),\\
\left.\frac{\partial \hat{\psi}_N(x)}{\partial x}\right|_{x=l}
-\left.\frac{\partial \hat{\psi}_R(x)}{\partial x}\right|_{x=l}
&=(Z\hat{I}-\lambda\hat{\tau})\hat{\psi}_N(l),
\end{align}
where $\hat{I}$ is the $4\times4$ identity matrix and
\begin{align}
\hat{\tau}=
\begin{pmatrix}
0&1&0&0\\
-1&0&0&0\\
0&0&0&1\\
0&0&-1&0
\end{pmatrix}.
\end{align}
By matching the wave-functions in the different regions by using the boundary conditions, we obtain a system of equations described as $\hat{\bm A}{\hat {\bm x}}=0$ where $\hat{\bm A}$ is a $16\times 16$ matrix and $\hat {\bm x}=(a_1,a_2,b_1,b_2,c_1,c_2,d_1,d_2,a_R,b_R,c_R,d_R,a_L,b_L,c_L,d_L)^T$.
Then, the Andreev level is determined by the condition det$(A)=0$.

Hereafter, we assume that $k_F=q_F\equiv k$ and $|h_L|=|h_R|\equiv h$, and fix the direction of the Zeeman field in the left superconductor $\theta_L=0$ for simplicity.
In the numerical calculations, we set $h=0.2\Delta$. Such a magnitude of the exchange splitting is experimentally well within reach using sub-Tesla magnetic fields \cite{MESERVEY1994173}.
%%%%%%%%%%%%%%%%%%%%%%%%%%%%%%%%%%%%%%%%%%%%%%%%%%%
%%%%%%%%%%%%%%%%%%%%%%%%%%%%%%%%%%%%%%%%%%%%%%%%%%%
%%%%%%%%%%%%%%%%%%%%%%%%%%%%%%%%%%%%%%%%%%%%%%%%%%%
%%%%%%%%%%%%%%%%%%%%%%%%%%%%%%%%%%%%%%%%%%%%%%%%%%%
\section{Andreev reflection and tunneling conductance}\label{sec_AR}
\begin{figure}[t]
\begin{center}
\includegraphics[width=8cm]{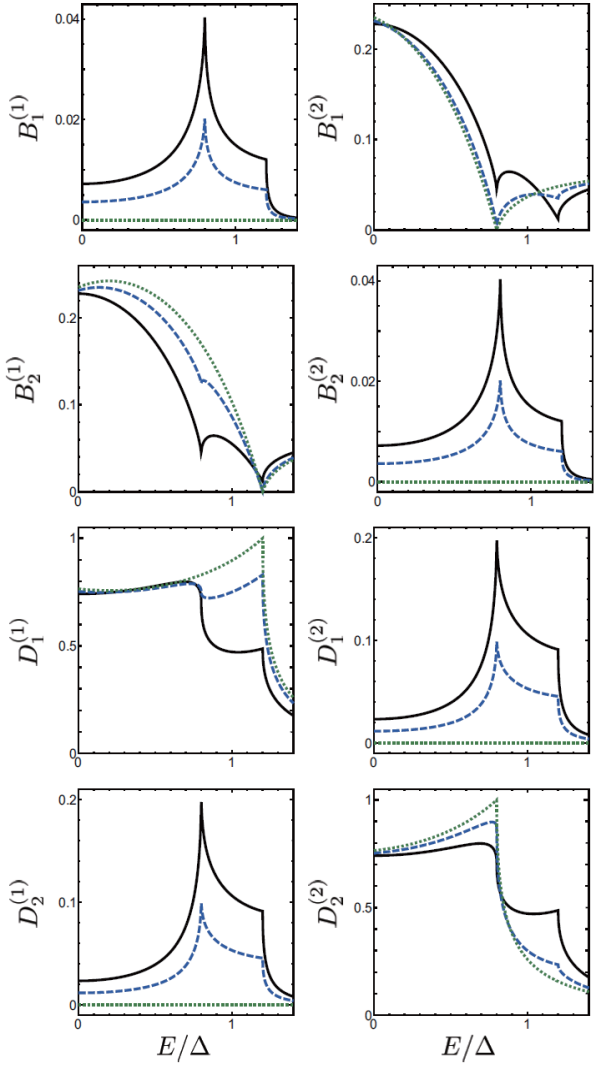}
\caption{Normal $B_j^{(i)}$ and Andreev $D_j^{(i)}$ reflection probability of the Rashba wire/Zeeman-split superconductor junction as a function of energy $E$ at $\zeta_R=\pi/2$ for $\theta_R=0$ (black solid line), $\theta_R=\pi/4$ (blue dashed line) and $\theta_R=\pi/2$ (green dotted line). The left (right) panels show the reflection probability for the electron with $k_1$ ($k_2$) injected case. Here, the parameters are set as $Z=0.5$ and $\lambda/k=1$.}
\label{fig_AR}
\end{center}
\end{figure}
We begin with the Andreev reflection process at the interface, since this process is of fundamental importance with regard to the formation of the Andreev levels we will later consider in a Josephson setup. Here, we consider only the right interface and set $l=0$. Then we obtain the reflection coefficients $b_1$, $b_2$, $d_1$ and $d_2$. 

In the presence of RSOI, the spin and momentum are locked and the right moving electron with wave number $k_1$ ($k_2$) is a spin-up (-down) eigenstate for the $y$ quantization axis in our model, which is shown in Fig.\ref{fig_setup} (b). Note that the left moving electron has the opposite spin compared with the right moving one. The reflection probability of a particle with $k_j$ ($j=1, 2$) for an electron with $k_i$ ($i=1, 2$) injected case is defined as $B_{j}^{(i)}=|b_j^{(i)}|^2$ and $D_{j}^{(i)}=|d_j^{(i)}|^2$. Here $B_{j}^{(i)}$ and $C_{j}^{(i)}$ are nomal and Andreev reflection probabilities, respectively. (See Appendix  for more information regarding $b_j^{(i)}$ and $d_j^{(i)}$)

In Fig.\ref{fig_AR}, we show numerical results of the normal and Andreev reflection probability at the interface for various orientations of the Zeeman field. The left (right) panels show the reflection probability for an electron with wave number $k_1$ ($k_2$) injected case. Let us start with the case where the orientation of the Zeeman field $\hat{\bm n}_R$ is parallel to the $z$ direction. In this case, the spin-dependent reflection has taken place at the interface, and thus all reflection probabilities for both $k_1$ and $k_2$ electron injected case are finite. Since the energy band in the superconducting region is lifted as shown in Fig.\ref{fig_setup} (c), there are double kink points at $E=\Delta \pm |h|$. As can be seen in Fig.\ref{fig_AR} (black line), the reflection probability for the electron with a $k_1$ injected case is fully consistent with that for the electron with $k_2$ injected case (though the spin is opposite), namely, 
$B_{1}^{(1)}=B_{2}^{(2)}$, $B_{2}^{(1)}=B_{1}^{(2)}$, $D_{1}^{(1)}=D_{2}^{(2)}$, and $D_{1}^{(2)}=D_{2}^{(1)}$.
With increasing $\theta_R$ for $\zeta_R=\pi/2$, $B_{1}^{(1)}$, $D_{2}^{(1)}$, $B_{2}^{(1)}$, and $D_{1}^{(2)}$ are suppressed, and when $\hat{\bm n}_R\parallel y$ ($\theta_R=\pi/2$), $B_{1}^{(1)}$, $D_{2}^{(1)}$, $B_{2}^{(2)}$ and $D_{1}^{(2)}$ become zero, since left- and right- moving particles with the same wave number have the opposite spin and there is no spin dependent scattering at the interface. The reflection probability $B_{2}^{(1)}$, $D_{1}^{(1)}$, $B_{1}^{(2)}$, and $D_{2}^{(2)}$ are also changed by tuning $\theta_R$ as shown in Fig.\ref{fig_AR}. At $\theta_R=\pi/2$, there is single kink point at $E=\Delta+h$ for $B_{2}^{(1)}$ and $D_{1}^{(1)}$, and at $E=\Delta-h$ for $B_{1}^{(2)}$ and $D_{2}^{(2)}$, since the spin-up and -down processes occurs separately. Note that here we vary the direction of the Zeeman field in the $z$-$y$ plane, but the same results can be obtained if we change the direction of the Zeeman field in the $x$-$y$ plane.
For more information, the analytical formulas for the reflection coefficients for $\hat{\bm n}_R\parallel z$ (or $x$) and $y$ are shown in Appendix A.

\begin{figure}[t]
\begin{center}
\includegraphics[width=8cm]{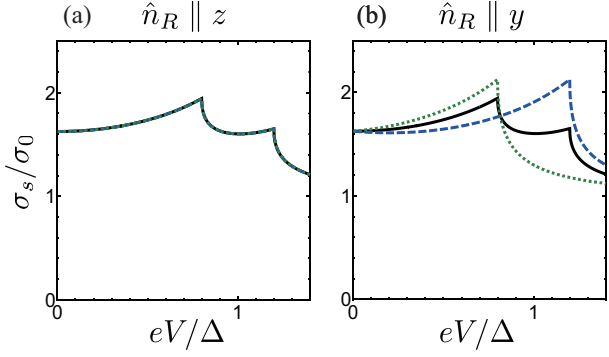}
\caption{Tunneling conductance of a Rashba wire/Zeeman-split superconductor junction as a function of bias voltage for (a) $\hat{\bm n}_R\parallel z$ and (b) $\hat{\bm n}_R\parallel y$. Black solid lines show the observable tunneling conductance. Blue dashed and green dotted lines are the contribution from the electron with $k_1$ and $k_2$ injected case, respectively. The total observable tunneling conductance does not depend on the orientation of the Zeeman field. }
\label{fig_didv}
\end{center}
\end{figure}

\begin{figure}[t]
\begin{center}
\includegraphics[width=8cm]{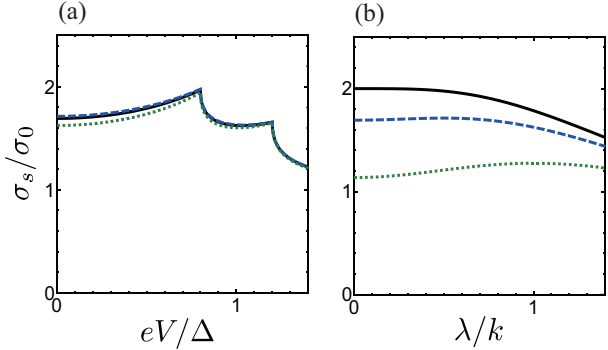}
\caption{(a) Tunneling conductance of Rashba wire/Zeeman split superconductor junction as a function of bias voltage at $Z=0.5$ for $\lambda/k=0$ (black solid line), $\lambda/k=0.5$ (blue dashed line) and $\lambda/k=1.0$ (green dotted line). (b) Tunneling conductance as a function of the strength of the RSOI $\lambda$ at zero voltage for $Z=0$ (black solid line), $Z=0.5$ (blue dashed line), and $Z=1.0$ (green dotted line).}
\label{fig_didv_lam_dep}
\end{center}
\end{figure}

%%%%%%%%%%%%%%%%%%%%%%%%%%%%%%%%%%%%%%%%%%%%%%%%%%%%%%%%%%%%%%%%%%%%
Next, we calculate the tunneling conductance $\sigma_s$ at zero temperature given by \cite{PhysRevB.25.4515,PhysRevB.74.035318,MIZUNO20091617}
\begin{align}
\sigma_s=\sum_{i=1,2} (1-B_{1}^{(i)}-B_{2}^{(i)}+D_{1}^{(i)}+D_{2}^{(i)}).
\end{align}
Figure \ref{fig_didv} (a) [(b)] shows the numerical results of the tunneling conductance in the case of $\hat{\bm n}_R\parallel z$ ($\hat{\bm n}_R\parallel y$). In these figures, the back solid lines show the observable tunneling conductance. Blue dashed and green dotted lines are the contribution from electrons with the $k_1$ and $k_2$ injected case, respectively. Although the contribution from the $k_1$ or $k_2$ injected case depends on the orientation of the Zeeman field when the fields are in the $z$-$y$ plane, the total observable tunneling conductance does not depend on the direction. This is because the contributions from the $k_1$ injected case and $k_2$ case completely compensate each other. 
Figure \ref{fig_didv_lam_dep} (a) shows the tunneling conductance as a function of the bias voltage for various $\lambda$, and Fig.\ref{fig_didv_lam_dep}(b) shows it as a function of $\lambda$ at zero bias voltage for various strengths of the barrier potential $Z$. As can be seen in the figures, the tunneling conductance is weakly affected by the RSOI, similarly to the results for two-dimensional electron gas with RSOI/superconductor junctions obtained by Yokoyama $et$. $al$ \cite{PhysRevB.74.035318}. Nevertheless, the Andreev levels in the ZS/RW/ZS trilayer are dramatically affected by the RSOI as shown below.


%%%%%%%%%%%%%%%%%%%%%%%%%%%%%%%%%%%%%%%%%%%%%%%%%%%
%%%%%%%%%%%%%%%%%%%%%%%%%%%%%%%%%%%%%%%%%%%%%%%%%%%
%%%%%%%%%%%%%%%%%%%%%%%%%%%%%%%%%%%%%%%%%%%%%%%%%%%
\section{Andreev levels in the absence of Rashba spin-orbit interaction}\label{sec_AL_noRSOI}
\begin{figure}[t]
\begin{center}
\includegraphics[width=8cm]{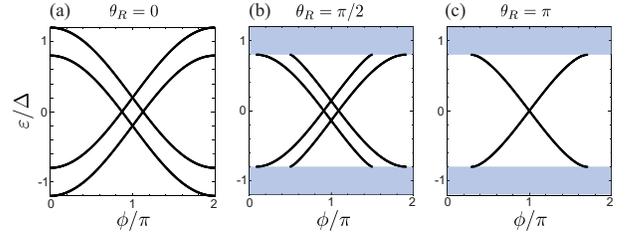}
\caption{Andreev level $\varepsilon$ as a function of the phase difference $\phi$ in the absence of the RSOI for (a) $\theta_R=0$, (b) $\theta_R=\pi/2$, and (c) $\theta_R=\pi$, where the direction of the Zeeman field in the left superconductor is fixed to $\theta_L=0$. When $\theta_L\neq 0$, the Andreev levels $\varepsilon$ for $E-\Delta<|\varepsilon|<E+\Delta$ are absent, the regions of which are shown with blue shading.}
\label{fig_noRSOI}
\end{center}
\end{figure}

Before we move to the main results of the Andreev level in the presence of RSOI, we briefly review the case without RSOI \cite{0295-5075-115-6-67001}. In the absence of the RSOI, the Andreev bound state changes depending on the relative direction of the Zeeman field in two superconductors. When the Zeeman fields in two superconductors are parallel, the spin degeneracy of the Andreev level is lifted and the Andreev level is described by 
\begin{align}
\varepsilon=\pm\Delta\cos \frac{\phi}{2} \pm h,
\label{eq_para}
\end{align}
where $\phi=\phi_R-\phi_L$ and $Z=0$. The first term of the right hand of Eq. (\ref{eq_para}) is the Andreev level in the absence of the Zeeman field \cite{Kulik}. Namely, the effect of the Zeeman field is simply the energy shift of $\pm h$ as shown in Fig.\ref{fig_noRSOI} (a). In this case, the Andreev level exists for $-\Delta-h<E<\Delta+h$. 
By changing the relative direction of the Zeeman field from the parallel configuration (vary $\theta_R$), the magnitude of the band splitting decreases.
When the Zeeman field in the right superconductor has a finite angle, the spin dependent Andreev reflection occurs at the right boundary, and a right-moving electron with up-spin is reflected as a left moving hole that has both spin up and down component. However, if the energy of the Andreev-reflected hole is less than $-\Delta+h$ (or the energy of the injected electron is more than $\Delta-h$), the spin-up component of the hole is not Andreev reflected at the left boundary because of the spin-splitting energy gap of the left superconductor. This means that the amplitude of the wave decays for every single scattering process and the Andreev bound state can not be formed for the energy region. Therefore, there are no lines for $\Delta-h<|E|<\Delta+h$ in Fig.\ref{fig_noRSOI} (b). When Zeeman fields in the two superconductors are antiparallel the energy band is degenerate, as shown Fig. \ref{fig_noRSOI} (c). 
In this case, the Andreev level is described by the following expression:
\begin{align}
\varepsilon=\pm\cot \frac{\phi}{2} \sqrt{\Delta^2(1-\cos \phi)-2h^2}.
\label{eq_al_ap}
\end{align}
It follows from Eq.(\ref{eq_al_ap}) that the Andreev level exists for $\phi_c<\phi<2\pi -\phi_c$ with $\phi_c=\cos^{-1}(1-2h/\Delta)$. 

%%%%%%%%%%%%%%%%%%%%%%%%%%%%%%%%%%%%%%%%%%%%%%%%%%%
%%%%%%%%%%%%%%%%%%%%%%%%%%%%%%%%%%%%%%%%%%%%%%%%%%%
%%%%%%%%%%%%%%%%%%%%%%%%%%%%%%%%%%%%%%%%%%%%%%%%%%%
\section{Andreev levels in the presence of Rashba spin-orbit interaction}\label{sec_AL_RSOI}
\begin{figure}[t]
\begin{center}
\includegraphics[width=8cm]{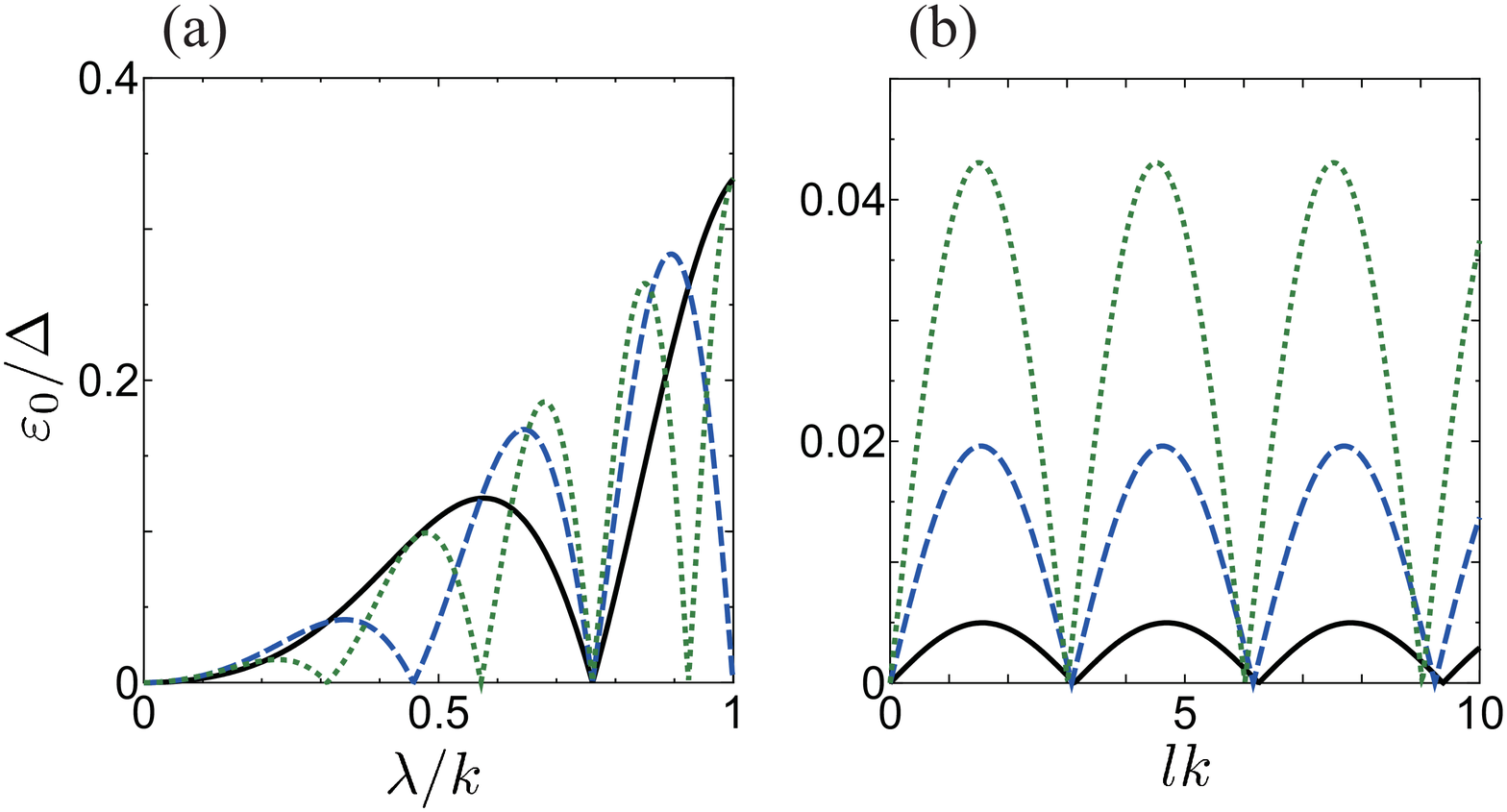}
\caption{(a) The magnitude of the energy gap $\varepsilon_0$ at $\phi=\pi$ as a function of $\lambda$ at $lk=10$ (black solid), $lk=20$ (blue dashed) and $lk=30$ (green dotted). (b) The magnitude of the energy gap at $\phi=\pi$ as a function of $l$ at $\lambda/ k=0.1$ (black solid), $\lambda/ k=0.2$ (blue dashed), and $\lambda/ k=0.3$ (green dotted).}
\label{fig_gapopen}
\end{center}
\end{figure}

We now turn to the case with finite RSOI, which has not been studied previously in the literature.
In the presence of RSOI, a degenerate energy band of the normal region splits into two branches and the wave vectors have a different value from that in the superconducting region. This wave vector mismatch causes a natural barrier at the interface and leads to the energy gap at $\phi=\pi$.
This gap opening effect at $\phi=\pi$ can be seen in the absence of the Zeeman field and does not depend on the relative direction of the Zeeman field and the RSOI.
The analytical expression of the Andreev level in the absence of the Zeeman field is given by
\begin{align}
\varepsilon=\pm\Delta\sqrt{\frac{1}{2}\left( 1+\frac{4k^2(k^2+\lambda^2)\cos\phi+\lambda^4\sin^2(\sqrt{k^2+\lambda^2}l)}
{4k^2(k^2+\lambda^2)-\lambda^4\sin^2(\sqrt{k^2+\lambda^2}l)} \right)},
\label{eq_al_h0}
\end{align} 
and the energy gap at $\phi=\pi$ is
\begin{align}
\varepsilon_0=\Delta\sqrt{\frac{\lambda^{4}\sin^2(\sqrt{k^2+\lambda^{2}}l)}
{4k^2(k^2+\lambda^{2})+\lambda^{4}\sin^2(\sqrt{k^2+\lambda^{2}}l)} },
\label{eq_ep0}
\end{align}
where we set $Z=0$ for simplicity.
As seen from the above equation, the magnitude of the energy gap depends on $\lambda$, $k$, and $l$. The energy gap is closed when the parameters satisfy $\sqrt{k^2+\lambda^{2}}l=n\pi$ where $n$ is an integer number.
Figure \ref{fig_gapopen} (a) shows the magnitude of the energy gap at $\phi=\pi$ as a function of $\lambda$ for various $l$. The magnitude of the energy gap increases with increasing $\lambda$ with oscillation and the energy gap is closed at 
\begin{align}
\lambda/k=\sqrt{(n\pi/kl)^2-1}. 
\end{align}
Figure \ref{fig_gapopen} (b) shows the magnitude of the energy gap as a function of $l$ for various $\lambda$. The magnitude of the energy gap oscillates by changing $l$, but the maximum value of each interval does not change. The energy gap is closed at 
\begin{align}
kl=n\pi/\sqrt{1+\lambda^{2}/k^2},
\end{align}
and the period of the oscillation slightly decreases with increasing $\lambda$. 
Note that, in the presence of the Zeeman field, the energy bands in the superconducting region also split so that the wave vectors for up spin and down spin are different. However, the difference between the wave vectors for up spin and down spin caused by the Zeeman field are much smaller than that caused by the RSOI since here we restrict the Zeeman energy $h<\Delta$. Therefore, we ignore the effect of the wave-vector mismatch originating in the superconducting region.

\begin{figure*}[t]
\begin{center}
\includegraphics[width=17cm]{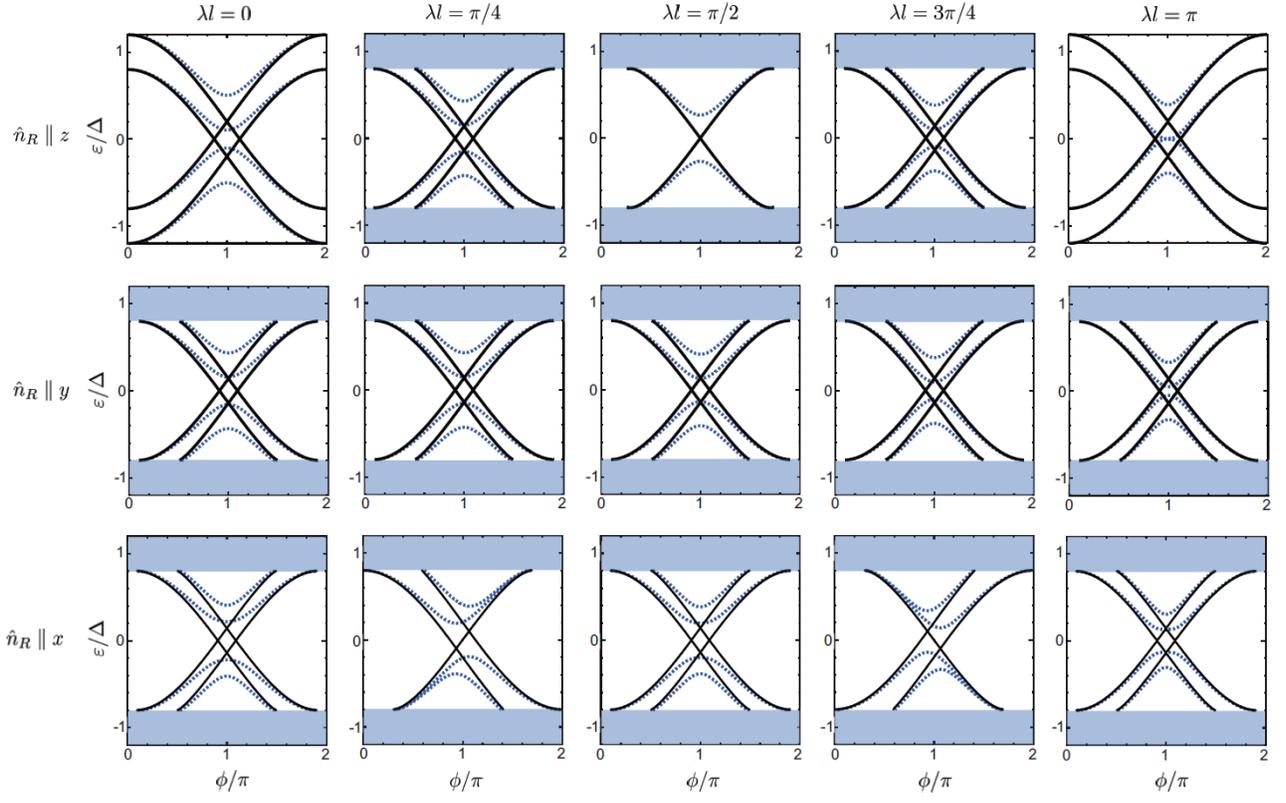}
\caption{Andreev levels $\varepsilon$ as a function of the phase difference $\phi$ for various $\lambda$ values in the direction of the Zeeman fields in the right superconductor ${\bm \hat{n}_R}\parallel z$ (upper low), ${\bm \hat{n}_R}\parallel y$ (middle low) and ${\bm \hat{n}_R}\parallel x$ (bottom low). Here, the direction of the Zeeman field in the left superconductor is fixed to $\hat{n}_L\parallel z$. From left to right, $\lambda/k$ varies from 0 to $\pi/kl$ by $\pi/4kl$. The black solid (blue dotted) lines show the Andreev levels in the case of $Z=0$ ($Z=0.5$).}
\label{fig_with_RSOI}
\end{center}
\end{figure*}

\begin{figure}[tb]
\begin{center}
\includegraphics[width=5cm]{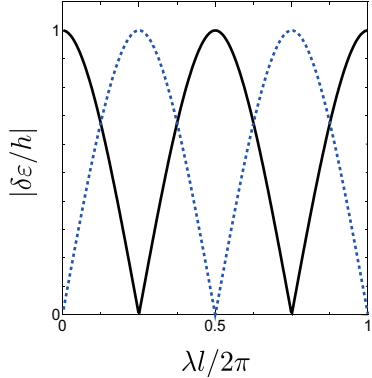}
\caption{Normalized magnitude of the energy shift $\delta\varepsilon/h$ as a function of $\lambda l/2\pi$ for the parallel (black solid line) and antiparallel (blue dotted) Zeeman case.}
\label{fig_gapde}
\end{center}
\end{figure}

Next, we discuss the phenomena that can be seen only in the simultaneous presence of the Zeeman field and RSOI. We begin with the case where the Zeeman field in both superconductors are oriented in the $+z$ direction. In Fig.\ref{fig_with_RSOI} (first row), we show the numerical results of the Andreev level $\varepsilon$ as a function of $\phi$ for various $\lambda$. Here, black solid (blue dotted) lines show the Andreev level for $Z=0$ ($Z=0.5$). As shown in the figure, the magnitude of the energy splitting changes depending on $\lambda$ and $l$, with a period $2\pi/\lambda l$. Note that in the numerical calculation, we have set $\lambda\ll k$, which gives rise to a quite small gap at $\phi=\pi$ in the figure. 
At $\lambda=n\pi/l$, the magnitude of the band splitting has a maximum value for any $\phi$. On the other hand, if $\lambda=(n+1/2)\pi/l$, the magnitude of the band splitting is minimum; $\delta\varepsilon=0$ for any $\phi$. The black solid line in Fig. \ref{fig_gapde} shows the magnitude of the band splitting at $\phi=\pi$ as a function of the $\lambda l /2\pi$. This oscillation of the band splitting $\delta\varepsilon$ can be described by 
\begin{align}
\delta\varepsilon=h\cos (\lambda l).
\label{eq_yzy_p}
\end{align}
Note that if the Zeeman field in the two superconductors is an antiparallel configuration, the Andreev level is degenerated in the absence of the RSOI. Then with increasing $\lambda$ or $l$, the magnitude of the band splitting oscillates, as shown by the blue dotted line in Fig. \ref{fig_gapde}. 
In the presence of the barrier potential, the gap $\phi=\pi$ is enhanced as shown in Fig. \ref{fig_with_RSOI}. On the other hand, the oscillation period of the magnitude of the band splitting is not affected by the barrier potential.

This oscillatory behavior can be understood physically by the spin precession that takes place in the Rashba wire. If an electron or hole traveling in the Rashba wire has a spin component perpendicular to the $y$ direction, the spin precession occurs \cite{doi:10.1063/1.102730,RSOI}. The precession angle is given by 
\begin{align}
\theta_P=(k_2-k_1)l=2\lambda l.
\end{align}
When $\theta_P=\pi$, a spin-up particle is converted to that with down-spin by traveling through the Rashba wire and vice versa. In this case, even if the Zeeman field in both superconductors is parallel ($+z$ direction), the particles behave as if the Zeeman field in the superconductors is antiparallel ($+z$ and $-z$ direction). As a result, the
 magnitude of the Andreev level is  the same as that in the case of an antiparallel Zeeman field without RSOI, which can be seen by comparing the third panel of the first row of Fig.\ref{fig_with_RSOI} and Fig.\ref{fig_noRSOI}(c). 
Not only for the parallel or antiparallel Zeeman case but also for the arbitrary $\theta_R$ with fixed $\zeta_R=\pi/2$, the magnitude of the energy splitting of the Andreev level is the same as that with $\theta_R=\theta_P$ for the non-RSOI case. Note that although the magnitudes of the band splitting are identical to each other, the shapes of the Andreev levels are not the same. This is because an energy gap appears at $\phi=\pi$ in the presence of RSOI.
 In addition, the spin precession is not affected by the barrier potential. Therefore the oscillation period of the band splitting is not changed even in the presence of the barrier potential. 

We also find that the oscillatory behavior of the band splitting by the RSOI changes depending on the relative direction of the Zeeman field in the two superconductors. First, we change the direction of the Zeeman field in the right superconductor in the $y$-$z$ plane, i.e., vary $\theta_R$ for $\zeta_R=\pi/2$. 
With increasing $\theta_R$ for fixed $\zeta_R=\pi/2$, the amplitude of the band splitting becomes smaller. 
Then if the Zeeman field in the right superconductor is parallel to the $y$ direction, the magnitude of the band splitting does not depend on $\lambda$ and $l$, as shown in the middle row of Fig. \ref{fig_with_RSOI}. Note that as long as the Zeeman field in either left or right superconductor is parallel to the $y$ direction, the magnitude of the band splitting does not depend on $\lambda$ and $l$. This is because, the spin of a particle coming from the superconductor with a $y$-oriented Zeeman field does not precess.

Next, we change the direction of the Zeeman field in the right superconductor in the $z$-$x$ plane.
If the Zeeman field of the right superconductor has the $x$ component, the $\varepsilon$-$\phi$ curve becomes asymmetric for $\phi$, $\varepsilon(\phi)\neq\varepsilon(2\pi-\phi)$, as shown in the third row of Fig.\ref{fig_with_RSOI}. In addition, the same as the parallel Zeeman case, the magnitude of the band splitting oscillates as a function of $\lambda$ and $l$ with a period $2\pi/\lambda l$. The asymmetric Andreev level and resulting anomalous Josephson effect are predicted to be realized in S/F/S junctions with spin active interfaces \cite{PhysRevB.76.224525, grein_prl_09, kulagina_prb_14}, S/F/S junctions with the RSOI in the normal region \cite{PhysRevLett.101.107001,PhysRevLett.101.107005,PhysRevB.89.195407,0953-8984-27-20-205301}, S/N/S junctions with unconventional superconductors \cite{0034-4885-63-10-202,doi:10.1143/JPSJ.69.1152} and many other systems, e.g., Refs.\cite{PhysRevLett.103.107002,PhysRevB.81.184525,PhysRevB.92.100503}. 
In most systems considered so far, to achieve the anomalous Josephson effect, it is necessary to manipulate the magnetic field, which is experimentally difficult. On the other hand, in our system, the anomalous Josephson effect can be realized by changing the strength of the RSOI, which is experimentally feasible.

There is another feature originating from the RSOI: the disappearance of the Andreev level for $\Delta-h<|E|<\Delta+h$. Even if one introduces a small magnitude of the RSOI and the Zeeman field is parallel, the Andreev level of the energy region suddenly disappears. This is because in the presence of RSOI, there is the spin-dependent scattering at the interface. This spin-dependent scattering prohibits the formation of the bound state, as is discussed for the case of without RSOI in Sec.\ref{sec_AL_noRSOI}. 

The main effects of the RSOI on the shape of the Andreev levels are (i) band shift, which is represented by Eq. (\ref{eq_yzy_p}), and (ii) gap opening effect at $\phi=\pi$, which is captured by Eq. (\ref{eq_al_h0}).
In the presence of both RSOI and Zeeman field, the analytical formula of the Andreev level is quite complicated. However, by combining the above effects represented by Eqs. (\ref{eq_al_h0}) and (\ref{eq_yzy_p}), we derive the following approximate solution for the parallel Zeeman case ($z$ direction):
\begin{align}
\varepsilon=&\pm\Delta\sqrt{\frac{1}{2}\left( 1+\frac{4k^2(k^2+\lambda^2)\cos\phi+\lambda^4\sin^2(\sqrt{k^2+\lambda^2}l)}
{4k^2(k^2+\lambda^2)-\lambda^4\sin^2(\sqrt{k^2+\lambda^2}l)} \right)}
\nonumber
\\
&\pm h \cos \lambda l.
\label{eq_i_try}
\end{align}
This solution reproduces the numerical results quite well especially near $\phi=\pi$.

Finally, we briefly comment on the difference between the Andreev levels in the present system and those in the S/F/S junctions. 
The periodic change of the Andreev level is also known for the S/F/S junction \cite{Chtchelkatchev2001,Kuplevakhskii,CAYSSOL200694}. 
In the S/F/S junction, the Andreev level is given by 
\begin{align}
\varepsilon=\pm\Delta\cos\frac{\phi\pm lk\rho}{2},
\end{align}
where $\rho$ is the magnitude of the exchange field normalized by the Fermi energy and  is considered as $\rho\ll 1$.
In this case, with increasing the strength of the exchange field, the degenerate branches shift to the $\pm\phi$ direction (horizontal direction in $\varepsilon$-$\phi$ plot), 
which causes the $\pi$ transition.
On the other hand, in the present system, the energy band shifts to a vertical direction by changing the strength of the RSOI, which does not cause a $\pi$ transition.

%%%%%%%%%%%%%%%%%%%%%%%%%%%%%%%%%%%%%%%%%%%%%%%%%%%%%%%%%%%%%
%%%%%%%%%%%%%%%%%%%%%%%%%%%%%%%%%%%%%%%%%%%%%%%%%%%%%%%%%%%%%
%%%%%%%%%%%%%%%%%%%%%%%%%%%%%%%%%%%%%%%%%%%%%%%%%%%%%%%%%%%%%
%%%%%%%%%%%%%%%%%%%%%%%%%%%%%%%%%%%%%%%%%%%%%%%%%%%%%%%%%%%%%
\section{Summary}\label{sec_summary}
In summary, we have theoretically studied how the Rashba spin-orbit interaction in the normal region and the Zeeman field in the superconducting region affect the formation of Andreev levels in a Josephson junction. 
We have found that the total tunneling conductance remains invariant, whereas the Andreev reflection processes and the resulting Andreev levels are sensitive to the relative orientation of the spin-orbit field and the Zeeman field. 

We have shown that the Andreev level is systematically changed by tuning the strength of the Rashba spin-orbit interaction $\lambda$ or length of the Rashba wire $l$. In particular, the magnitude of the band splitting $\delta\varepsilon$ oscillates as a function of $\lambda$ and $l$, and we have clarified that this behavior is interpreted physically by the spin precession in the Rashba wire. It has been also found that the $\varepsilon$-$\phi$ curve changes depending on the relative angle of the three independent vectors, i.e., the orientation of Zeeman fields in the left superconductor, that in the right superconductor, and the vector characterizing the spin-orbit interaction. In particular, the $\varepsilon$-$\phi$ curve becomes asymmetric with respect to the phase difference $\phi$ when either the left or right Zeeman field has a component parallel to the junction ($x$ component).
An interesting future direction is the possibility to control the Josephson current in the 
considered system by the change of the Andreev levels, which will be a subject of future study. 
Moreover, it would be also interesting to study the finite frequency response of this system as discussed in other systems  \cite{PhysRevB.87.174521,PhysRevB.92.134508}.

\acknowledgments

Y. Asano, A. Br$\o$yn, and A. Brinkman are thanked for useful discussions. T.H. was supported in part by a Grant in Aid for JSPS Fellows (No. 26010542) and is thankful for the kind hospitality of NTNU where the part of this work was performed. A.A.G. was supported by JSPS-RFBR Bilateral Joint Research Projects and Seminars Grant No. 17-52-50080,
Russian Science Foundation Grant No. 15-12-30030, Russian-Greek project No.  RFMEFI61717X0001, and 
the Ministry of Education and Science of the Russian Federation Grant No. 14.Y26.31.0007.
Y.T. was supported by a Grant-in-Aid for Scientific
Research on Innovative Areas Topological Material Science
JPSJ KAKENHI (Grants No. JP15H05851, No. JP15H05853,
and No. JP15K21717), a Grant-in-Aid for Scientific Research
B (Grant No. JP15H03686), and a Grant-in-Aid for
Challenging Exploratory Research (Grant No. JP15K13498)
from the Ministry of Education, Culture, Sports, Science, and
Technology, Japan (MEXT). 
J.L. was supported by funding via the “Outstanding Academic Fellows” programme at NTNU, the COST Action MP-1201, the NV-Faculty, the Research Council of Norway Grants No. 216700 and No. 240806. This work was partly supported by the Research Council of Norway through its Centres of Excellence funding scheme, project “QuSpin”.

\appendix

\section{Andreev reflection coefficients}
In this Appendix, we show the analytical formulas for the normal and Andreev reflection coefficients of the RM/ZS junction. The reflection coefficients in the case of $\hat{\bm n}_R\parallel z$ are given by
\begin{align}
b_{1}^{(1)}&=[2kk_\lambda \lambda^2 (u_2v_1-u_1v_2)(u_2v_1+u_1v_2)]/\Gamma,\\
b_{2}^{(1)}&=\lambda^2[(k^2+k_{\lambda}^2)(u_1^2-v_1^2)(u_2^2-v_2^2)
\nonumber\\
&+2k k_\lambda (u_1^2u_2^2-v_1^2v_2^2)]/\Gamma,\\
d_{1}^{(1)}&=2ikk_\lambda(u_2 v_1+u_1v_2)[(k^2+k_{\lambda}^2)(u_1 u_2- v_1 v_2)\nonumber\\
&+2kk_\lambda(u_1u_2+v_1v_2)]/\Gamma,\\
d_{2}^{(1)}&=2ikk_\lambda(u_2 v_1-u_1v_2)[(k^2+k_{\lambda}^2)(u_1 u_2+ v_1 v_2)
\nonumber\\
&+2kk_\lambda(u_1u_2-v_1v_2)]/\Gamma,\\
b_{1}^{(2)}&=b_{2}^{(1)},
b_{2}^{(2)}=b_{1}^{(1)},
d_{1}^{(2)}=d_{2}^{(1)},
d_{2}^{(2)}=d_{1}^{(1)},
\end{align}
with
\begin{align}
k_\lambda&=\sqrt{k^2+\lambda^2},\\
\Gamma&=[(k^2+k_{\lambda}^2)(u_1^2-v_1^2)+2kk_\lambda(u_1^2+v_1^2)]\nonumber\\
&\times[(k^2+k_{\lambda}^2)(u_2^2-v_2^2)+2kk_\lambda(u_2^2+v_2^2)].
\end{align}
Here, $b_{1}^{i}$, $b_{2}^{i}$, $d_{1}^{i}$, and $d_{2}^{i}$ with $i=1(i=2)$ are the reflection coefficients $b_{1}$, $b_{2}$, $d_{1}$, and $d_{2}$ in Eq.(\ref{eq_wave_RW}) when an electron with $k_1$ ($k_2$) is injected. On the other hand, the reflection coefficients in the case of $\hat{\bm n}_R\parallel y$ are given by
\begin{align}
b_{1}^{(1)}&=0,\\
b_{2}^{(1)}&=\frac{\lambda^2(u_1^2-v_1^2)}{(k^2+k_{\lambda}^2)(u_1^2-v_1^2)+2kk_\lambda(u_1^2+v_1^2)},\\
d_{1}^{(1)}&=\frac{4ikk_\lambda u_1 v_1}{(k^2+k_{\lambda}^2)(u_1^2-v_1^2)+2kk_\lambda(u_1^2+v_1^2)},\\
d_{2}^{(1)}&=0,\\
b_{1}^{(2)}&=\frac{\lambda^2(u_2^2-v_2^2)}{(k^2+k_{\lambda}^2)(u_2^2-v_2^2)+2kk_\lambda(u_2^2+v_2^2)},\\
b_{2}^{(2)}&=0,\\
d_{1}^{(2)}&=0,\\
d_{2}^{(2)}&=\frac{4ikk_\lambda u_2 v_2}{(k^2+k_{\lambda}^2)(u_2^2-v_2^2)+2kk_\lambda(u_2^2+v_2^2)}.
\end{align}

%%%%%%%%%%%%%%%%%%%%%%%%%%%%%%%%%%%%%%%%%%%%%%%%%%%%%%%%%%%
%%%%%%%%%%%%%%%%%%%%%%%%%%%%%%%%%%%%%%%%%%%%%%%%%%%%%%%%%%%
%%%%%%%%%%%%%%%%%%%%%%%%%%%%%%%%%%%%%%%%%%%%%%%%%%%%%%%%%%%

\bibliography{ref}

\end{document}